\documentclass[a4paper,11pt]{article}
\pdfoutput=1
\usepackage[T1]{fontenc}
\usepackage{amsmath,amssymb,url}
\usepackage{graphicx}
\usepackage{cite,hyperref}
\usepackage{subcaption}
\usepackage{multirow}
\usepackage{xspace}
\providecommand{\href}[2]{#2}

\newcommand\as{\alpha_{\mathrm{S}}}

\def\to{\rightarrow}
 
\def\nn{\nonumber}

\newcommand{\Hh}{\ensuremath{H}}
\newcommand{\HH}{\ensuremath{HH}}
\newcommand{\HHj}{\ensuremath{HH+\mathrm{jet}}}
\newcommand{\pT}{\ensuremath{p_{\mathrm{T}}}\xspace}
\newcommand{\pTHone}{\ensuremath{p_{\mathrm{T},H_1}}\xspace}
\newcommand{\pTHtwo}{\ensuremath{p_{\mathrm{T},H_2}}\xspace}
\newcommand{\pTHH}{\ensuremath{p_{\mathrm{T},HH}}\xspace}
\newcommand{\mHH}{\ensuremath{m_{HH}}\xspace}

\newcommand{\pTj}{\ensuremath{p_{\mathrm{T},j_1}}\xspace}
\newcommand{\sqrtS}{\ensuremath{\sqrt{s}}}
\newcommand{\ord}{\mathcal{O}}

\newcommand\Matrix{{\sc Matrix}\xspace}
\newcommand\Munich{{\sc Munich}\xspace}
\newcommand\OpenLoops{{\sc OpenLoops}\xspace}

\newcommand\Sherpa{{\sc Sherpa}\xspace}
\newcommand\powheg{{\sc POWHEG-BOX}\xspace}
\newcommand\mcfm{{\sc MCFM}\xspace}
\newcommand{\LHAPDF}{{\rmfamily\scshape Lhapdf}\xspace}
\newcommand{\CutTools}{{\sc CutTools}\xspace}
\newcommand{\OneLOop}{{\sc OneLOop}\xspace}
\newcommand{\qT}{q_{\mathrm{T}}}

\newcommand{\qTHH}{q_{\mathrm{T},\HH}}

\newcommand{\D}{\mathrm{d}}

\newcommand{\kT}{k_{\mathrm{T}}}
\newcommand{\nf}{n_F}

\newcommand{\pph}{\ensuremath{pp \to H+X}}
\newcommand{\pphh}{\ensuremath{pp \to HH+X}}
\newcommand{\pphhj}{\ensuremath{pp \to \HHj+X}}
\newcommand{\abbrev}{}

\newcommand{\lo}{\text{\abbrev LO}\xspace}
\newcommand{\nlo}{\text{\abbrev NLO}\xspace}
\newcommand{\nnlo}{\text{\abbrev NNLO}\xspace}

\newcommand{\qcd}{{\abbrev QCD}\xspace}
\newcommand{\rF}{\mathrm{F}}
\newcommand{\rR}{\mathrm{R}}
\newcommand{\GeV}{\text{GeV}\xspace}

\newcommand{\beqar}{\begin{eqnarray}}
\newcommand{\eeqar}{\end{eqnarray}}
\newcommand{\beq}{\begin{equation}}
\newcommand{\eeq}{\end{equation}}
\newcommand{\bit}{\begin{itemize}}
\newcommand{\eit}{\end{itemize}}

\def\citere#1{\mbox{Ref.~\cite{#1}}}
\def\citeres#1{\mbox{Refs.~\cite{#1}}}
\def\reffi#1{\mbox{Fig.~\ref{#1}}}

\def\reffis#1#2{\mbox{Figs.~\ref{#1}--\ref{#2}}}
\def\refta#1{\mbox{Tab.~\ref{#1}}}

\def\refse#1{\mbox{Sect.~\ref{#1}}}

\def\refeq#1{\mbox{Eq.~(\ref{#1})}}

\newcommand\Tstrut{\rule{0pt}{3.0ex}}         
\newcommand\Bstrut{\rule[-1.5ex]{0pt}{0pt}}   

\def\relplotwidth{0.47}

\usepackage{pstricks}
\usepackage{graphicx}

\begin{document} 
\begin{titlepage}
\renewcommand{\thefootnote}{\fnsymbol{footnote}}
\begin{flushright}
DESY 16-107\\
FR-PHENO-2016-007\\
ICAS 08/16 \\
MITP/16-061 \\
ZU-TH 20/16
\end{flushright}
\vspace*{1.5cm}

\begin{center}
{\Large \bf Differential Higgs Boson Pair Production at\\[0.3cm] Next-to-Next-to-Leading Order in \qcd }
\end{center}


\par \vspace{2mm}
\begin{center}
{\bf Daniel de Florian$^{(a)}$},
{\bf Massimiliano Grazzini$^{(b)}$},
{\bf Catalin Hanga$^{(b)}$},
{\bf Stefan Kallweit$^{(c)}$}, \\[0.1cm]
{\bf Jonas M. Lindert$^{(b)}$}, 
{\bf Philipp Maierh\"ofer$^{(d)}$},
{\bf Javier Mazzitelli$^{(a)}$}, 
{\bf Dirk Rathlev$^{(e)}$}
\vspace{5mm}

$^{(a)}$ International Center for Advanced Studies (ICAS), UNSAM, Campus Miguelete \\
25 de Mayo y Francia, (1650) Buenos Aires, Argentina\\[0.3cm]
$^{(b)}$ Physik-Institut, Universit\"at Z\"urich, \\ 
Winterthurerstrasse 190, CH-8057 Z\"urich, Switzerland \\[0.3cm]
$^{(c)}$ PRISMA Cluster of Excellence, Institute of Physics,\\
Johannes Gutenberg University, D-55099 Mainz, Germany\\[0.3cm]
$^{(d)}$ Physikalisches Institut, Albert-Ludwigs-Universit\"at Freiburg,\\
79104 Freiburg, Germany\\[0.3cm]
$^{(e)}$ Theory Group, Deutsches Elektronen-Synchrotron, D-22607 Hamburg, Germany

\vspace{5mm}

\end{center}

\par \vspace{-0.5cm}
\begin{center} {\large \bf Abstract} \end{center}
\begin{quote}
\pretolerance 10000

We report on the first fully differential calculation for double Higgs boson
production through gluon fusion in hadron collisions up to
next-to-next-to-leading order (NNLO) in QCD perturbation theory. The calculation
is performed in the heavy-top limit of the Standard Model, and in the
phenomenological results we focus on $pp$ collisions at $\sqrtS=14$\,TeV. We
present differential distributions through NNLO for various observables
including the transverse-momentum and rapidity distributions of the two Higgs
bosons. NNLO corrections are at the level of $10\%-25\%$ with respect to the
next-to-leading order (NLO) prediction with a residual scale uncertainty of
$5\%-15\%$ and an overall mild phase-space dependence. Only at NNLO the
perturbative expansion starts to converge yielding overlapping scale uncertainty
bands between NNLO and NLO in most of the phase-space.
The calculation includes NLO predictions for \pphhj{}. Corrections to the 
corresponding distributions exceed $50\%$ with a residual scale dependence of 
$20\%-30\%$.


\end{quote}

\vspace*{\fill}
\begin{flushleft}

\end{flushleft}
\end{titlepage}
\setcounter{footnote}{1}
\renewcommand{\thefootnote}{\fnsymbol{footnote}}


\section{Introduction}
\label{sec:intro}

The discovery of the Higgs boson \cite{Aad:2012tfa,Chatrchyan:2012xdj} 
during Run I of the Large Hadron Collider (LHC) opened the door towards direct
tests of electroweak symmetry breaking. To this end the search for the production 
of Higgs boson pairs is one of the main goals of ongoing and future runs of the
LHC. Only this production mode allows for direct tests of Higgs trilinear self-couplings, 
whose knowledge in turn is necessary to reconstruct the scalar potential responsible for
electroweak symmetry breaking.

As it is the case for single Higgs boson production, the dominant production mode for 
Higgs boson pairs in the Standard Model (SM) proceeds at hadron colliders via 
gluon fusion, mediated by heavy-quark loops. 
At the leading order (LO) in QCD~\cite{Glover:1987nx,Eboli:1987dy,Plehn:1996wb} there 
are two interfering production mechanisms: either the two Higgs bosons
couple directly to a heavy-quark loop via a box diagram, $gg \to HH$, or they couple
via the trilinear Higgs coupling $\lambda$ to an off-shell Higgs boson, which in turn is 
produced via a triangular loop, similarly to single Higgs boson production, $gg \to H^*\to HH$.

Due to the loop suppression of the LO process and additional large accidental 
cancellations between ``triangle'' and ``box'' contributions in the scattering 
amplitude~\cite{Glover:1987nx}, signal rates for Higgs boson pair production are only 
at the level of few fb at $8$\,TeV and of tens of fb at $13-14$\,TeV.
These small rates, together 
with large irreducible backgrounds in the relevant $b \bar b \gamma 
\gamma$~\cite{Baglio:2012np,Yao:2013ika,Barger:2013jfa}, $b \bar b \tau 
\bar\tau$~\cite{Yao:2013ika,Barger:2013jfa,Dolan:2012rv}, $b \bar b W^+ 
W^-$~\cite{Papaefstathiou:2012qe} and $b \bar b b \bar 
b$~\cite{deLima:2014dta,Behr:2015oqq} final states, pose a true challenge to 
experimental searches for Higgs boson pair production. Consequently, currently only 
upper limits exist, constraining the cross section for Higgs boson pair production to 
the level of about $70$ times the SM 
prediction~\cite{Aad:2015uka,Aad:2014yja,Aad:2015xja,Khachatryan:2015yea,CMS:2014ipa,Khachatryan:2016sey}. 
However, the production cross section can significantly be 
altered by new-physics effects, for example due to new loop 
contributions~\cite{Dawson:2015oha}, altered $t\bar t h$ or novel $t\bar t h h$ 
couplings~\cite{Contino:2012xk}, or due to new resonances~\cite{Barger:2014taa}. Thus, future measurements 
(together with precision predictions) on Higgs boson pair production, and in particular
ratios of cross-section measurements~\cite{Goertz:2013kp}, do not just serve as stringent 
tests of electroweak symmetry breaking in the SM, 
but might also open the door towards physics beyond the SM. 

Given the loop-induced nature of the scattering process for Higgs boson pair 
production, higher orders in perturbation theory are extremely difficult to 
calculate. Only very recently a complete next-to-leading order (NLO) calculation
became available~\cite{Borowka:2016ehy}, where the required multi-scale
two-loop scattering amplitudes have been evaluated via numerical integration. 
Prior to the pioneering work of \citere{Borowka:2016ehy} the possible impact of the
NLO corrections has been studied in 
\citeres{Grigo:2014jma,Grigo:2014oqa,Grigo:2013rya,Grigo:2015dia,Degrassi:2016vss}
via asymptotic expansions of the two-loop virtual diagrams in the inverse top-quark mass. 

Assuming the top quark to be heavy and all other quarks to be massless, 
an effective theory can be formulated, introducing a tree-level coupling 
of gluons and Higgs bosons. In this heavy-top limit NLO corrections to 
Higgs boson pair production have been presented in~\citere{Dawson:1998py}, where a
rescaling with the exact Born cross section was performed.
The obtained NLO corrections in the so-called Born-improved heavy-top limit
increase the total cross section by about $100\%$ with sizable 
remaining scale uncertainties. In contrast, the exact NLO computation presented 
in~\citere{Borowka:2016ehy} yields a result that is smaller by $14\%$.

In order to further improve on these predictions, and in particular to
reduce the remaining scale uncertainties, the effective theory allows 
for the computation of perturbative corrections beyond NLO.
To this end, employing the amplitudes derived in~\citere{deFlorian:2013uza} (supplemented 
by \citere{Grigo:2014jma}) a calculation of Higgs boson pair production at next-to-next-to-leading 
order~(NNLO) accuracy in the heavy-top approximation was presented in \citere{deFlorian:2013jea}. 
At the inclusive level the NNLO corrections increase the cross section 
by about $20\%$ with respect to the NLO prediction, leaving scale uncertainties 
at the level of $8\%-10\%$. 
Besides inclusive NNLO cross sections the calculation of 
\citere{deFlorian:2013jea} offers differential predictions in the invariant mass 
of the produced Higgs boson pair indicating a rather mild phase-space dependence of 
the NNLO corrections.
Still working in the heavy-top limit,
soft-gluon resummation up to next-to-next-to-leading logarithmic (NNLL) accuracy
has been carried out, and the results were matched to the NLO \cite{Shao:2013bz}
and the NNLO \cite{deFlorian:2015moa} fixed-order computations.
At NNLL+NNLO accuracy the theoretical uncertainties on the inclusive cross section due to 
QCD effects are reduced to about $5\%$~\cite{deFlorian:2015moa}.
Furthermore, extending the SM with additional dimension-6 operators~\cite{Goertz:2014qta}, 
NLO corrections in the heavy-top limit have been presented in~\citere{Grober:2015cwa}.
Notably, in \citeres{Maltoni:2014eza,Frederix:2014hta} a reweighting technique 
has been presented, allowing to combine exact one-loop real corrections of Higgs boson pair
production with the corresponding virtual contributions, the latter computed in the effective theory.
On the other hand, the real corrections, which imply the evaluation of one-loop 
amplitudes with an extra parton in the final state, have been computed in an 
exact way and used to merge LO samples for $HH+0,1$ 
jets~\cite{Li:2013flc,Maierhofer:2013sha} in order to obtain more reliable 
exclusive distributions.

In this paper we extend the calculation of~\citere{deFlorian:2013jea} providing 
fully differential NNLO predictions for Higgs boson pair production in the heavy-top
approximation of the SM via a flexible Monte Carlo implementation. The calculation
is based on the combination of the $\qT$ subtraction formalism~\cite{Catani:2007vq} with the
Monte Carlo framework \Munich{}\footnote{{\sc Munich} is the abbreviation of
``MUlti-chaNnel Integrator at Swiss (CH) precision''---an automated parton level 
NLO generator by S.~Kallweit. In preparation.},
supplemented by tree and one-loop amplitudes from \OpenLoops~\cite{OLhepforge}.
Employing these tools we provide NNLO predictions
for various kinematic distributions that are relevant for searches and precision measurements of
Higgs boson pair production at the LHC. The calculation includes NLO predictions for \pphhj{}.
Corresponding differential distributions are studied in detail. 
In our study we refrain from a reweighting using exact LO or NLO  matrix elements or 
cross sections. Such a reweighting should eventually be performed 
employing the results of~\citere{Borowka:2016ehy}.
For the time being we focus on the differential NNLO/NLO correction factors obtained in 
the effective theory, which are the main result of our paper. Such correction factors
provide valuable information that can directly be applied to any \pphh{} NLO prediction 
in different approximations.

The paper is organized as follows. In \refse{sec:ingredients} we introduce the
heavy-top limit for multi-Higgs production at higher orders in perturbation
theory together with the technical ingredients of our calculation.
Numerical results are presented in \refse{sec:results}, and in \refse{sec:summary}
we summarize our results.

\section{Technical ingredients}
\label{sec:ingredients}

\subsection{Higgs boson pair production through NNLO in the heavy-top limit}
In the heavy-top approximation effective tree-level couplings between gluons and Higgs bosons are introduced via the 
effective Lagrangian~\cite{Djouadi:1991tka,Spira:1995rr,Dawson:1998py}
\beq\label{eq:LHEFT}
\mathcal{L}_{\text{HEFT}} = - \frac{1}{4} G_{\mu \nu} G^{\mu \nu} \left(C_H \frac{H}{v} - C_{HH} \frac{H^2}{v^2} \right)\, ,
\eeq
where $v\simeq 246\,\GeV$ is the vacuum expectation value of the Higgs field. In this effective Lagrangian
only couplings relevant for our calculation are shown, while in general within this effective theory 
there are also further couplings for any number of Higgs bosons to gluons. The matching coefficients 
$C_{\Hh}$ and $C_{\HH}$ can be expanded in powers of $\as$ via the following parametrization,
\beq
C_{X} = -\frac{\as}{3\pi}\sum_{n\geq 0} C_{X}^{(n)}
\left(\frac{\as}{\pi}\right)^n\, , \text{ with }
X=\Hh,\HH \, .
\eeq
The perturbative expansion for both coefficients is known up to  $\ord(\as^3)$ and 
reads \cite{Kramer:1996iq,Chetyrkin:1997iv,Djouadi:1991tka,Grigo:2014jma} 
\beqar
\label{eq:match}
C_{\Hh}^{(0)} &=& C_{\HH}^{(0)} \;\;=\;\; 1 \,, \\
C_{\Hh}^{(1)} &=& C_{\HH}^{(1)} \;\;=\;\; \frac{11}{4} \,, \nn\\
C_{\Hh}^{(2)} &=& \frac{2777}{288} + \frac{19}{16}\ln \frac{\mu_R^2}{m_t^2} + \nf \left(-\frac{67}{96}+\frac{1}{3}\ln \frac{\mu_R^2}{m_t^2} \right) \,, \nn\\
C_{\HH}^{(2)} &=& C_{\Hh}^{(2)} + \frac{35}{24} + \frac{2 \nf}{3}\,,\nn
\eeqar
where $\nf$ is the number of light quarks, $\mu_R$ the renormalisation scale and $m_t$ the pole mass 
of the (heavy) top quark. As can be seen from \refeq{eq:match}, up to $\ord(\as^2)$ we have $C_{\Hh} = C_{\HH}$. 

As already discussed, in the full theory Higgs boson pair production at LO is governed 
by ``box'' and ``triangle'' contributions.
In the heavy-top limit the corresponding scattering amplitudes 
manifest as tree-level diagrams with one double-Higgs and one single-Higgs operator 
insertion, respectively.
At NLO, in the perturbative expansion we have the usual real and virtual contributions, 
where the former includes gluon and quark bremsstrahlung, and the latter are given by one-loop 
corrections to the diagrams mentioned before.
However, at the same order of perturbation theory there is an additional 
contribution with Born-level kinematics, originating from amplitudes with two 
single-Higgs operator insertions in interference with the LO 
amplitude~\cite{Dawson:1998py}. In the full theory such contributions correspond 
to reducible double-triangle two-loop diagrams.

This pattern also appears at higher orders, and in particular the NNLO virtual contributions 
have to include both two-loop corrections to amplitudes with one single- or 
double-Higgs operator insertion, and one-loop corrections to amplitudes with two 
single-Higgs operator insertions~\cite{deFlorian:2013uza}. These NNLO virtual contributions
have to be combined via an appropriate subtraction scheme with double-real and 
real--virtual contributions of the same perturbative order.
Similarly to what was discussed before,
the real--virtual contributions, i.e. the virtual amplitudes 
for $HH+\text{jet}$ production, have to be extended to include two single-Higgs 
operator insertions in interference with the corresponding tree-level amplitude.
More details on the technical implementation of such double-operator 
insertions in our calculation are given in \refse{sec:openloops}.

\subsection{$\mathbf{\qT}$ subtraction}

In order to handle infrared singularities in the NNLO calculation, we apply the $\qT$ subtraction 
formalism~\cite{Catani:2007vq}. In this approach the separation between genuine NNLO singularities, 
located where the transverse momentum of the Higgs pair, $\qTHH$,  is zero, 
from NLO-like singularities in the $\HHj$ contribution is explicit. As a consequence, the contribution 
$\D{\sigma}^{\mathrm{HH + jet}}_{\mathrm{NLO}}$ in the $\qT$ subtraction formula,
 \begin{align}
\label{eq:main}
\D{\sigma}^{\HH}_{\mathrm{NNLO}}={\cal H}^{\HH}_{\mathrm{NNLO}}\otimes \D{\sigma}^{\HH}_{\mathrm{LO}}+\left[ \D{\sigma}^{\HHj}_{\mathrm{NLO}}-\D{\sigma}^{\mathrm{CT}}_{\mathrm{NNLO}}\right],
\end{align}
can be evaluated using any well-established subtraction procedure at NLO. The remaining 
divergence in the limit $\qTHH\to0$ is cancelled by the process-independent counterterm 
$\D{\sigma}^{\mathrm{CT}}_{\mathrm{NNLO}}$. The implementation is fully general, 
and it is based on the universality  \cite{Catani:2013tia} of
the hard-collinear coefficients ${\cal H}^{\HH}_{\mathrm{NNLO}}$
appearing in the first term on the right hand side of Eq.~(\ref{eq:main}). The general structure
of these coefficients
for gluon-initiated processes has been presented in \citeres{Catani:2011kr,Gehrmann:2014yya}.
Their process dependence is embodied in a single perturbative hard factor
which is obtained from the two-loop virtual correction, derived for this process
in \citere{deFlorian:2013uza},
through an appropriate subtraction procedure \cite{Catani:2013tia}.

\begin{figure}[t]
\begin{center}
\includegraphics[width=\relplotwidth\textwidth]{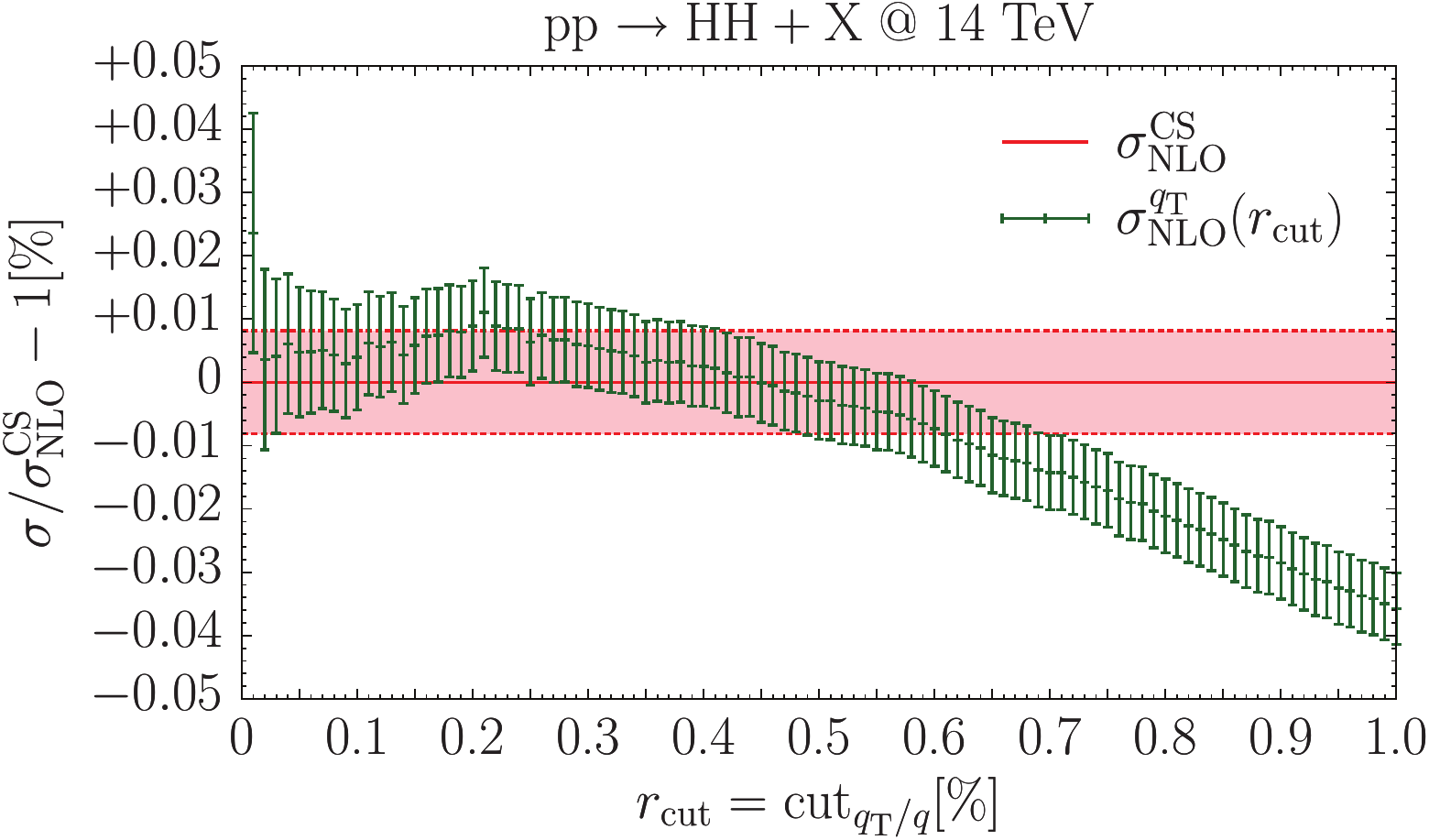}\hfill
\includegraphics[width=\relplotwidth\textwidth]{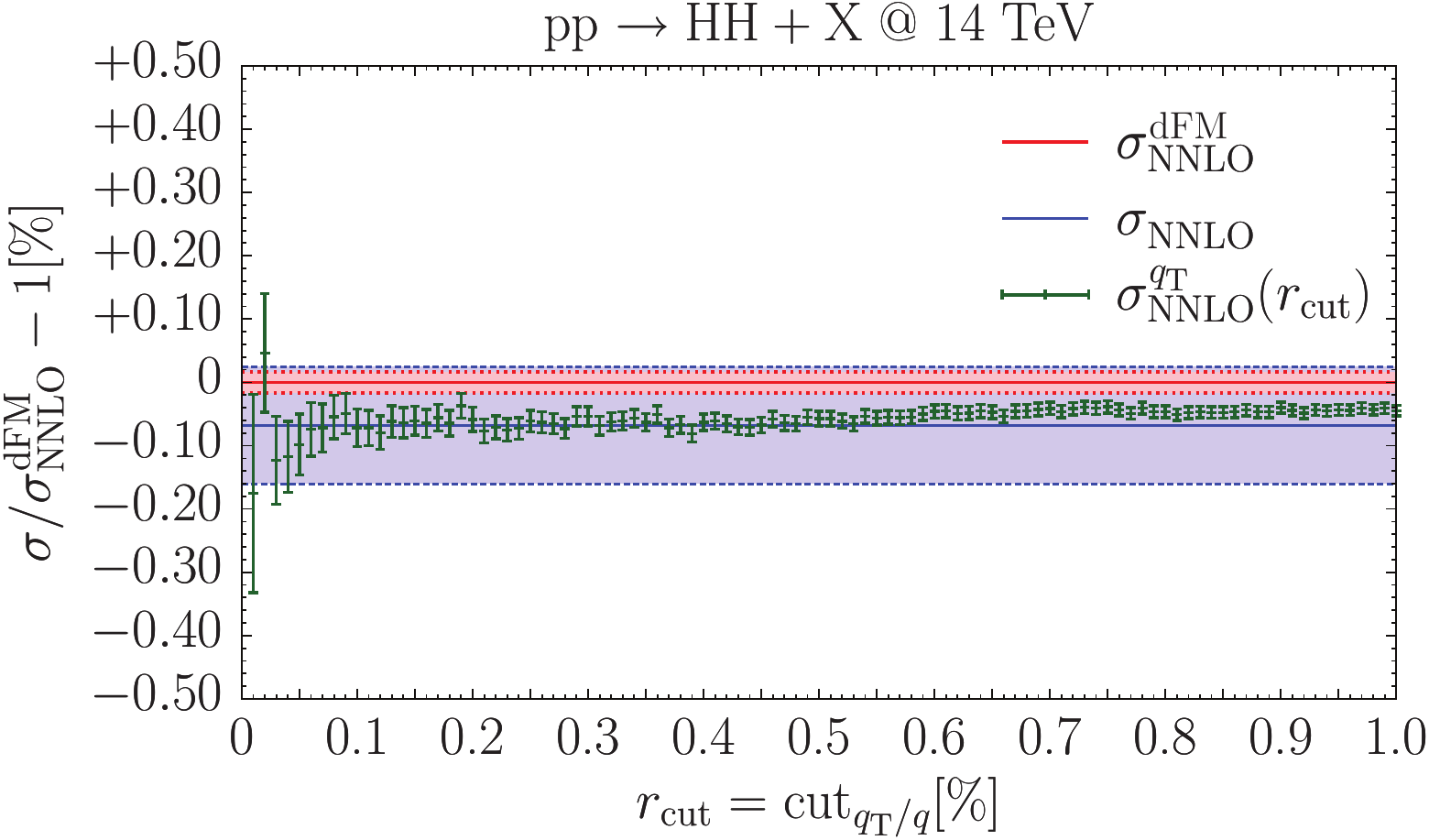}\\[1ex]
\caption[]{\label{fig:stability}{
Dependence of the \pphh{} cross sections at 14\,TeV on the $\qT$-subtraction cut, $r_{\mathrm{cut}}$,
for both NLO (left plot) and NNLO (right plot) results. 
NLO results are normalized to the $r_{\mathrm{cut}}$-independent NLO cross section computed with Catani--Seymour subtraction, 
and the NNLO results are normalized to the $r_{\mathrm{cut}}$-independent inclusive NNLO cross section 
calculated in the framework of \citere{deFlorian:2015moa}. The blue band indicates the NNLO result 
from $\qT$ subtraction in the limit $r_{\mathrm{cut}}\to0$, with an approriate extrapolation-error estimate.
}
}
\end{center}
\end{figure}

The difference in the square bracket in Eq.~(\ref{eq:main}) is formally finite as $\qT\to 0$,
but each term separately exhibits logarithmic divergences in this limit.
In practice
a small technical cut, $r_{\mathrm{cut}}$, needs to be applied on $r\equiv \qT/Q$, where 
$Q$ is typically chosen as the invariant mass of the final-state system (so here 
$Q=\mHH$). 
After cancellation of these logarithms 
between the real contribution $\D{\sigma}^{\mathrm{HH + jet}}_{\mathrm{NLO}}$ and the counterterm,
the remainder shows a very slight $r_{\mathrm{cut}}$ dependence below about $r_{\mathrm{cut}}=1\%$;
we thus use the finite-$r_{\mathrm{cut}}$ results to extrapolate to $r_{\mathrm{cut}}=0$, 
and conservatively assign an additional extrapolation error to our results.
We verified in detail that for Higgs boson pair production the NNLO result is 
indeed very stable when varying the cut parameter. More precisely,
in the range below $r_{\mathrm{cut}}=1\%$, the variation in the NNLO cross section is 
of ${\cal O}(0.1\%)$, and our extrapolated result is in good numerical agreement
with the analytic result of \citere{deFlorian:2015moa} (see \reffi{fig:stability}).

\subsection{Tree and one-loop amplitudes from \textbf{\textsc{OpenLoops}}}
\label{sec:openloops}

All tree and one-loop amplitudes, i.e. in particular the one-loop amplitude for
the $\D\sigma^{\HHj}_\mathrm{NLO}$ contribution, are provided by the publicly
available\footnote{The publicly available \OpenLoops process library
includes all relevant matrix elements to compute NLO \qcd corrections, including
colour- and helicity-correlations and real radiation as well as loop-squared
amplitudes, for more than a hundred LHC processes. Amplitudes for Higgs boson pair
production (+1 jet) at NLO in the heavy-top limit have been made
available together with this publication, while amplitudes for $pp\to \Hh, pp\to
\Hh j,pp\to \Hh jj$ and $pp\to \Hh jjj$ have been publicly available already for
some time.}
\OpenLoops{} amplitude generator~\cite{OLhepforge}, which is based on a fast
numerical recursion for the generation of tree and one-loop scattering
amplitudes~\cite{Cascioli:2011va}.

In order to extend \OpenLoops to one-loop corrections in the heavy-top
limit of the SM, all relevant Feynman rules for single-Higgs~\cite{Kauffman:1996ix}
and double-Higgs~\cite{Kniehl:1995tn,Grober:2015cwa} production have been
implemented in the framework of the numerical open-loops recursion including UV
renormalisation and the rational contributions of type
$R_2$~\cite{Page:2013xla}. Combined with the OPP reduction
method~\cite{Ossola:2006us} implemented in \CutTools~\cite{Ossola:2007ax} and
the scalar one-loop library \OneLOop~\cite{vanHameren:2010cp} the employed
recursion permits to achieve very high CPU performance and a high degree of
numerical stability. The small fraction of numerically unstable matrix elements
is automatically detected and rescued through re-evaluation in quadruple
precision.

Technically, the effective field theory of \refeq{eq:LHEFT} introduces various 
features which do not appear in the Standard Model. Most notably, the Feynman rules for the
dimension-5 and -6 operators $G_{\mu\nu}G^{\mu \nu}\Hh(\Hh)$ introduce, apart
from 5- and 6-point vertices, the Lorentz structure $p_1 p_2\,g^{\mu\nu}-p_1^\nu
p_2^\mu$ (where $p_1^\mu$ and $p_2^\nu$ are the gluon momenta) which, if present in a
loop, raises the tensor rank of the amplitude by $2$. In the calculation of
$HH$(+jet) production at one-loop level such operators enter only once, leading
to a tensor rank up to one higher than the number of loop propagators. 
The reduction of such amplitudes with one ``additional'' tensor rank is supported by \CutTools{}.
Furthermore, the $\ord(\as)$ contributions to the matching coefficients $C_H$ and
$C_{HH}$ must be included. Considering the order of coupling powers it is natural to
treat these contributions as counterterms. As discussed above, at the same
order of perturbation theory as the one-loop scattering amplitudes for
$HH$(+jet) production, contributions from two single-Higgs operator
insertions at tree-level in interference with the LO tree-level amplitude with
one double-Higgs operator insertion have to be considered. Similarly to the
$\ord(\as)$ contributions to $C_H$ and $C_{HH}$, these contributions are
included via dedicated $\ord(\as)$ pseudo-counterterms.

\subsection{Validation}

One-loop amplitudes for single Higgs boson production plus up to two jets have 
been extensively validated against the results of 
\citeres{Berger:2006sh,Badger:2006us,Badger:2007si,Glover:2008ffa,Badger:2009hw,
Dixon:2009uk,Badger:2009vh} implemented in \Sherpa~\cite{Gleisberg:2008ta} and 
\mcfm~\cite{Campbell:2012am} (via the corresponding implementation in the \powheg 
\cite{Alioli:2008tz,Campbell:2012am}).
Due to the lack of publicly available alternatives, the validation of the 
one-loop amplitudes for Higgs boson pair production plus jets had to rely on various 
internal cross checks. 

We performed a calculation of \pph{} up to NNLO in the heavy-top limit 
in the same framework as employed for \pphh{}, and compared against the results obtained 
with the inclusive analytical codes of \citeres{Catani:2003zt,Anastasiou:2012hx,Harlander:2012pb},
where agreement well beyond the per mill level was found.
Due to the similarity of the two processes this serves as a strong cross check for many technical 
ingredients of the calculation presented here.

In order to validate all ingredients of the computation of Higgs boson pair production 
in the heavy-top limit presented in this paper, the \lo, \nlo and \nnlo 
inclusive cross sections computed in \citere{deFlorian:2013jea} have been reproduced 
at the per mill level\footnote{In \cite{deFlorian:2013jea}
the relation $C_{H}^{(2)}=C^{(2)}_{HH}$ was assumed due to the lack of knowledge of $C_{HH}$
up to $\ord(\alpha_S^3)$ at that time, while for this cross check the 
matching coefficients listed in \refeq{eq:match} have been applied in both calculations.} (see \reffi{fig:stability}).
Additionally, mutual agreement has been found for the invariant mass 
distribution $\mHH$ up to \nnlo comparing against the results of \citere{deFlorian:2013jea}.
Furthermore, the NLO results have been computed in the $\qT$ subtraction formalism 
and also employing the dipole subtraction framework~\cite{Catani:1996jh,Catani:1996vz} within \Munich, where 
again we found mutual agreement far beyond the per mill level (see again \reffi{fig:stability}).

\section{Results}
\label{sec:results}

In the following we present predictions for Higgs boson pair production at the LHC including perturbative 
fixed-order corrections up to NNLO in the heavy-top limit. Inclusive results will be presented 
for centre-of-mass energies of $\sqrtS=13$\,TeV and $\sqrtS=14$\,TeV, while at the differential level 
we restrict ourselves to $\sqrtS=14$\,TeV.
SM input parameters are chosen according to the recommendations of \cite{Denner:2047636}, which 
in particular implies $v=246.2$\,GeV, $m_t=173.2$\,GeV and
\beq
m_{H} = 125\,\GeV\,.
\eeq
Here, the top-quark mass does only enter via the NNLO contributions to the matching coefficients, 
as given in \refeq{eq:match}.
For the calculation of hadron-level cross sections we employ the PDF4LHC15~\cite{Butterworth:2015oua}
parton distribution functions~(PDFs), and use the corresponding NLO PDFs 
for our LO and NLO predictions and NNLO PDFs for the NNLO predictions.\footnote{To be precise, we use
the \texttt{PDF4LHC\_nlo\_30} and \texttt{PDF4LHC\_nnlo\_30} sets, interfaced through 
the \LHAPDF library \cite{Buckley:2014ana}.}
Couplings are evaluated using the running strong coupling provided by the respective PDFs.
All light quarks, including bottom quarks, are treated as massless particles, i.e. $\nf=5$, 
while the top quark does not contribute explicitly in the employed heavy-top limit.
To define jets, we employ the anti-$\kT$ jet clustering algorithm \cite{Cacciari:2008gp} with 
$R=0.4$ and require $p^{\mathrm{T}}_{j} > 30$\,GeV and $|\eta_j| < 4.4$.
In all results the renormalisation scale $\mu_\rR$ and factorisation scale $\mu_\rF$ are set to
\beq\label{eq:RFscales} 
\mu_{\rR,\rF}=\xi_{\rR,\rF}\mu_0,
\quad\mbox{with}\quad 
\mu_0= \mHH/2
\quad\mbox{and}\quad 
\frac{1}{2}\le \xi_{\rR},\xi_{\rF}\le 2 \, ,
\eeq 
where $\mHH$ is the invariant mass of the produced Higgs boson pair. Our default 
scale choice corresponds to $\xi_{\rR}=\xi_{\rF}=1$,
and theoretical uncertainties are assessed by
applying the 7-point scale variations $(\xi_\rR,\xi_\rF)=(2,2)$,
$(2,1)$, $(1,2)$, $(1,1)$, $(1,0.5)$, $(0.5,1)$, $(0.5,0.5)$, i.e.\ omitting 
antipodal variations. As shown in \citere{deFlorian:2015moa} the scale 
choice of \refeq{eq:RFscales} guarantees a good perturbative convergence of the 
total cross section and of the $\mHH$ distribution in Higgs boson pair production.

{\renewcommand{\arraystretch}{1.6}
\begin{table}[t]
\begin{center}
\begin{tabular}{|c|c|c|c|}
\hline
$\sqrtS$ [TeV] & $\sigma_{\lo}$ [fb]  & $\sigma_{\nlo}$  [fb]   & $\sigma_{\nnlo}$ [fb] 
\\[0.5ex]
\hline
\Tstrut
$13$           
& $13.8059(13)\,^{+31.5\%}_{-22.5\%}$ 
& $25.829(3)\,^{+17.8\%}_{-15.4\%}$ 
& $30.38(3)\,^{+5.2\%}_{-7.7\%}$     
\Bstrut\\
\Tstrut
$14$           
& $17.0778(16)\,^{+30.7\%}_{-22.1\%}$ 
& $31.934(3)\,^{+17.5\%}_{-15.1\%}$ 
& $37.52(4)\,^{+5.2\%}_{-7.6\%}$
\Bstrut\\[0.5ex]
\hline
\end{tabular}
\end{center}
\caption{Inclusive cross sections for Higgs boson pair production for different centre-of-mass energies
at LO, NLO and NNLO. 
Numerical errors on the respective previous digits are stated in brackets, including the
extrapolation error in the NNLO prediction.
Scale uncertainties are obtained from independent variations of $\mu_\rR$ and
$\mu_\rF$ around the central scale $\mu_0=\mHH/2$.
}
\label{tab:incl}
\end{table}

In \refta{tab:incl} we report inclusive cross sections for $\sqrtS=13$\,TeV and
$\sqrtS=14$\,TeV. No phase-space cuts are applied, and the quoted uncertainties
are obtained from scale variations. Both at $\sqrtS=13$\,TeV and $14$\,TeV the
NLO corrections increase the LO result by about $85\%$, and the NNLO corrections
have an effect of about $18\%$ on top of the NLO result. Scale uncertainties are
successively reduced from about $20\%-30\%$ at LO (which largely underestimates
the effect of higher-order corrections) to less than $10\%$ at NNLO.

\begin{figure*}[t!]
\centering
   \includegraphics[width=\relplotwidth\textwidth]{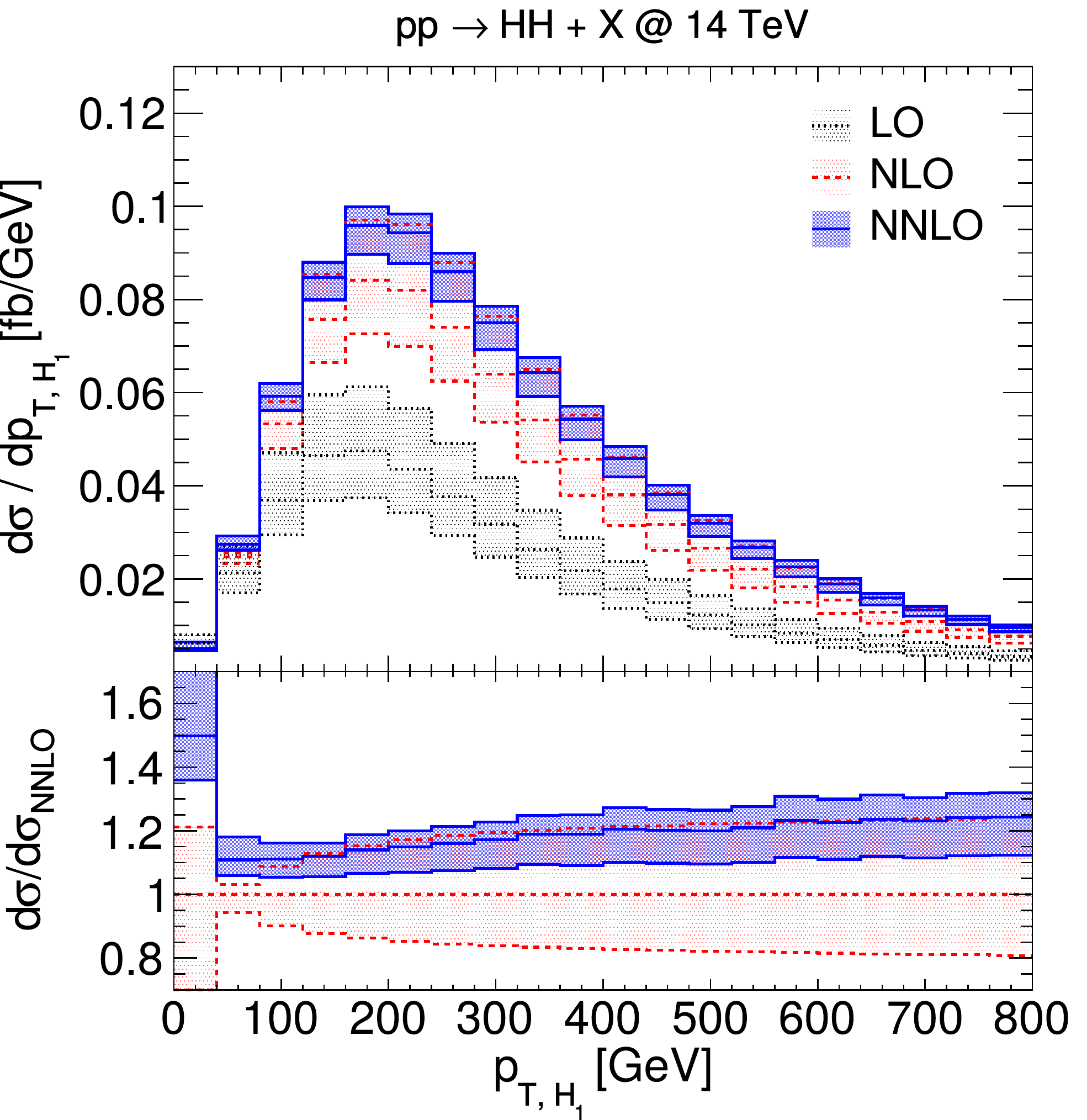}
         \qquad 
   \includegraphics[width=\relplotwidth\textwidth]{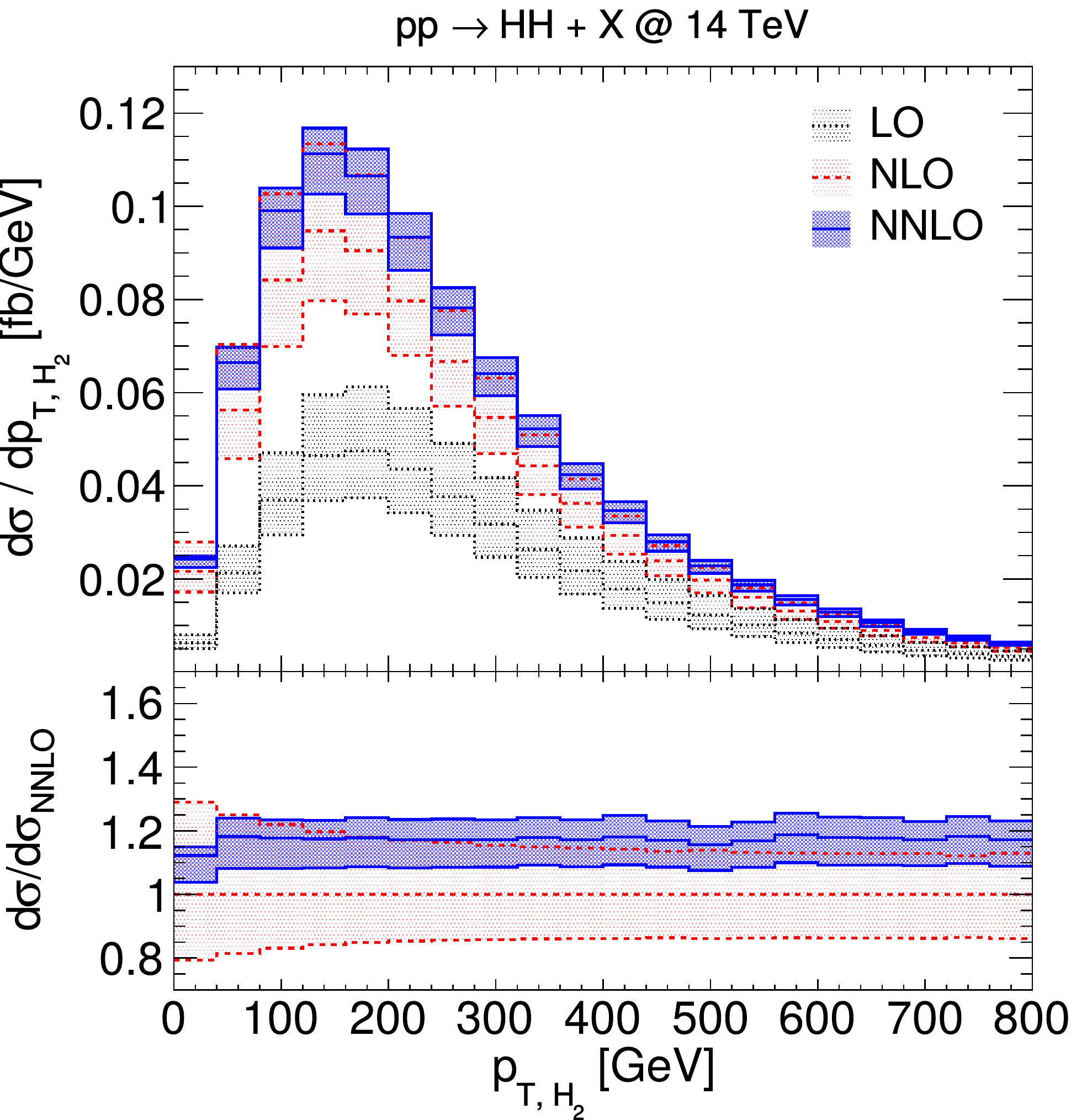}
\caption{
          Distributions in the transverse momentum of the harder (left) and the softer (right) 
          Higgs boson in \pphh{} at $\sqrtS=14$\,TeV.
          Shown are absolute LO (black), NLO (red) and NNLO (blue) predictions 
          in the heavy-top approximation and corresponding relative corrections  
          normalized to the central NLO prediction. Bands correspond to independent 
          variations of $\mu_\rR$ and $\mu_\rF$ around the central scale $\mu_0=\mHH/2$ as 
          described in the text.
          }
\label{fig:pTH1pTH2}
\end{figure*}

In \reffis{fig:pTH1pTH2}{fig:dRHj} differential distributions for 
Higgs boson pair production at the LHC with $\sqrtS=14$\,TeV are shown at LO, NLO and 
NNLO accuracy. In those distributions shown in 
\reffis{fig:pTH1pTH2}{fig:mHHyHH}, both NLO and NNLO corrections are 
sizable, and only at NNLO the perturbative convergence becomes manifest with 
overlapping scale uncertainty bands between the NLO and NNLO predictions in most 
of the considered phase-space regions. At the same time theoretical uncertainties estimated
by the scale variations described above are approximately halved when going from NLO to NNLO.
The NNLO distributions shown in \reffis{fig:pTHHpTj1}{fig:dRHj} are 
effectively only of next-to-leading order as they are either trivial or not 
defined at LO. They can be considered as a computation of $\HHj$ at 
NLO. Nevertheless, in the following discussion we always denote the highest 
considered perturbative order as NNLO (with respect to \pphh{}).

\begin{figure*}[t!]
\centering
   \includegraphics[width=\relplotwidth\textwidth]{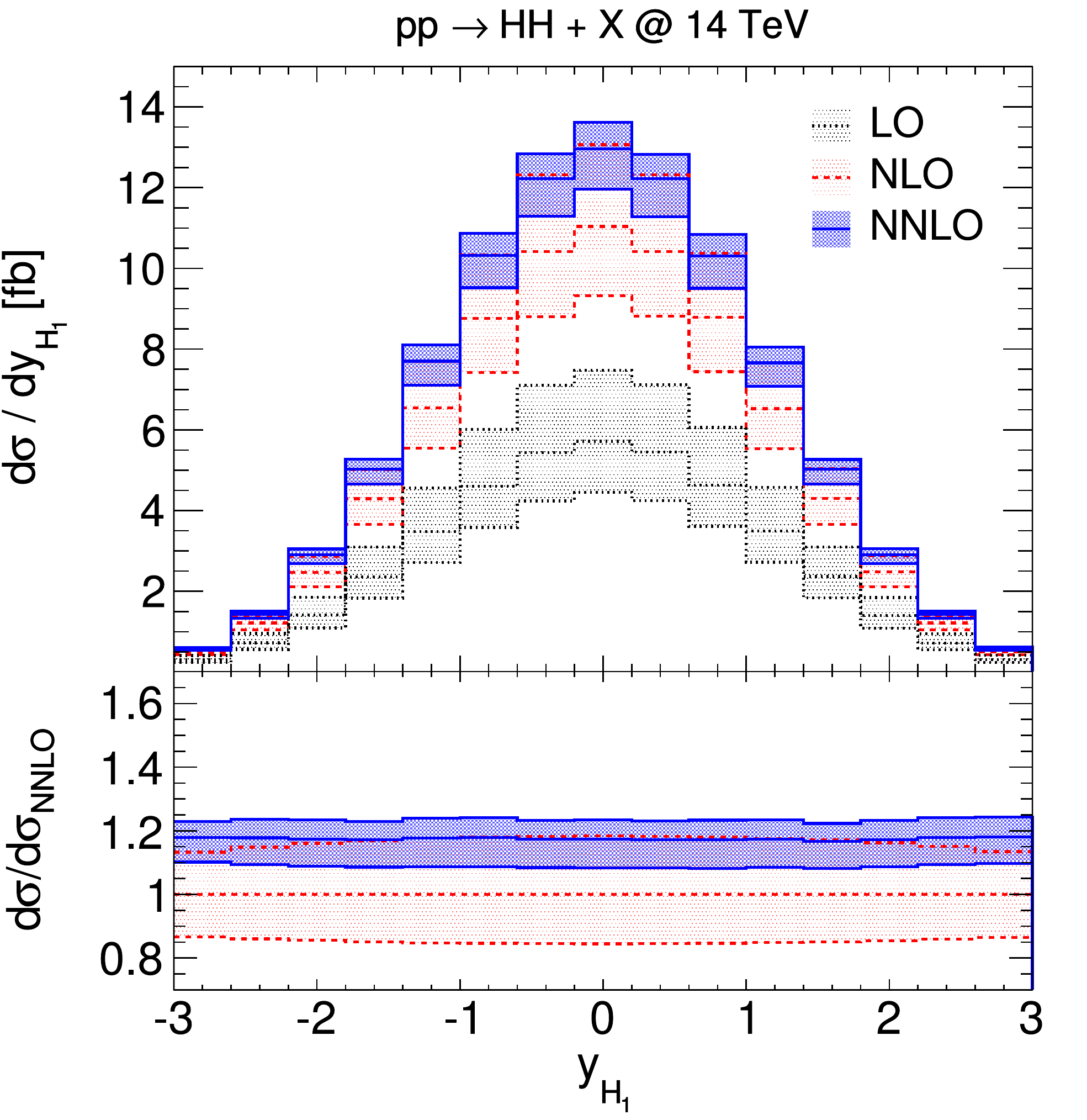}
         \qquad 
   \includegraphics[width=\relplotwidth\textwidth]{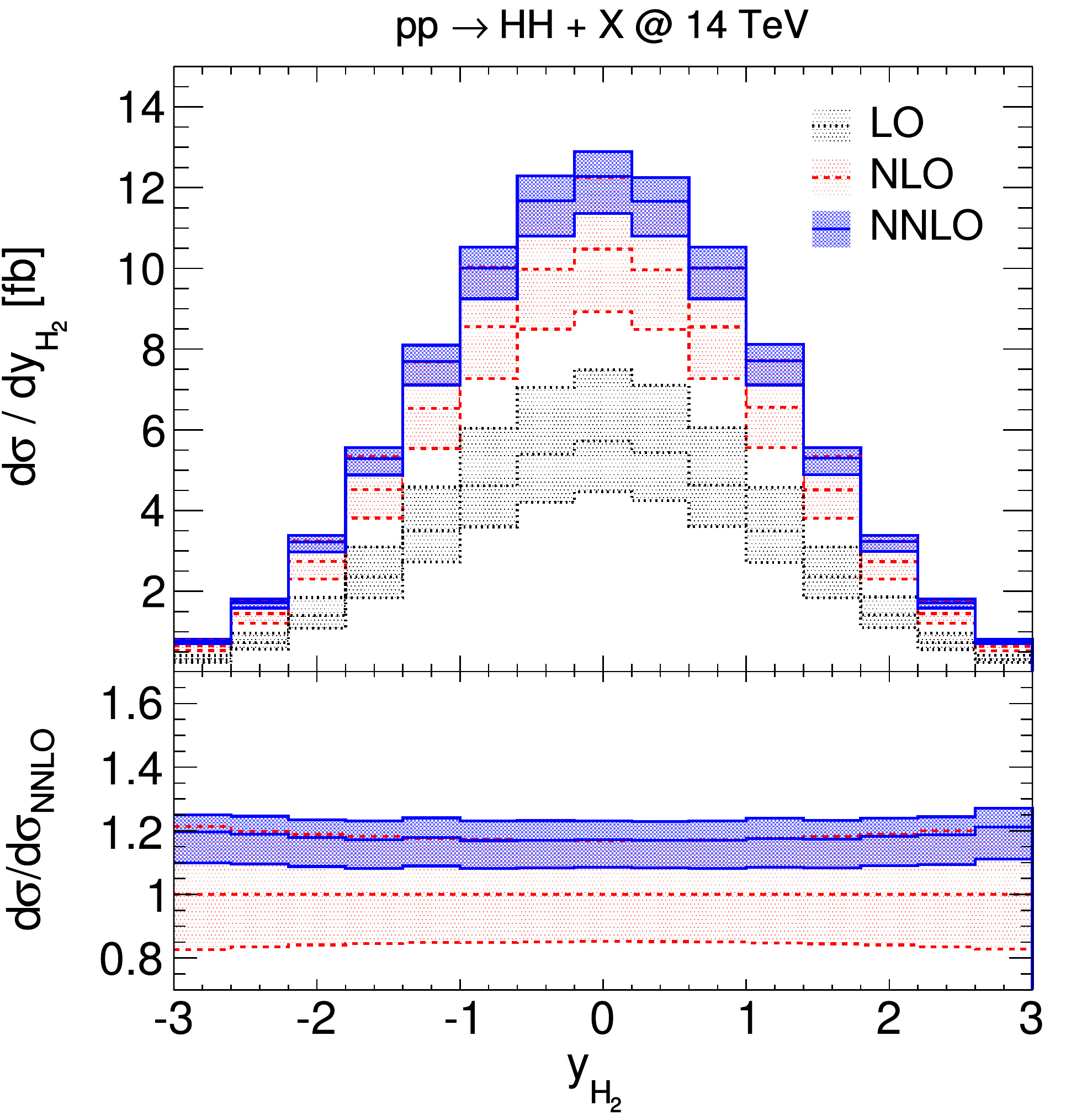}
\caption{
	  Distributions in the rapidity of the harder (left) and the softer (right) Higgs boson.
	  Curves and bands as in \reffi{fig:pTH1pTH2}.
	}
\label{fig:yH1yH2}
\end{figure*}

\begin{figure*}[t!]
\centering
   \includegraphics[width=\relplotwidth\textwidth]{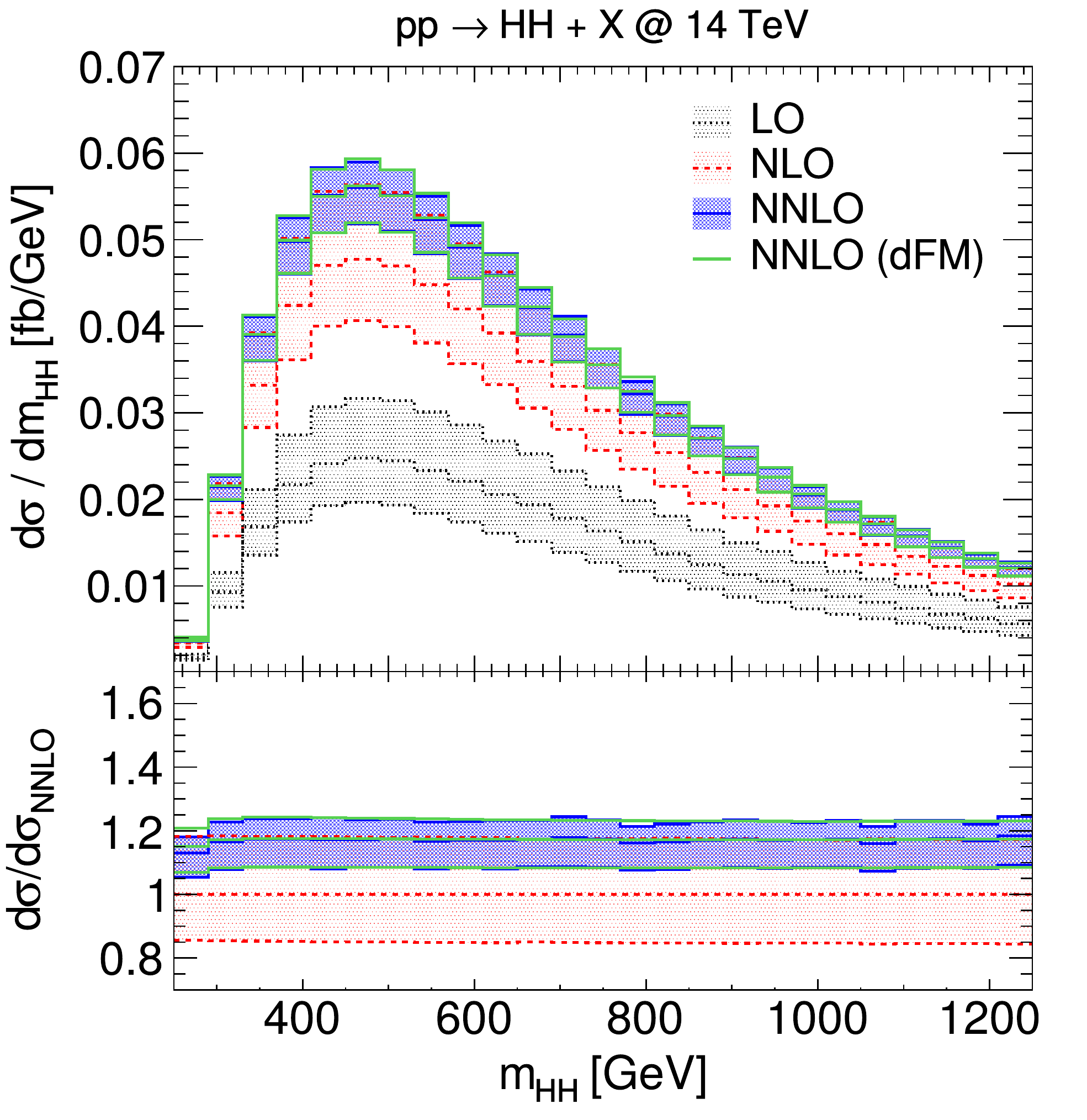}
   \qquad
  \includegraphics[width=\relplotwidth\textwidth]{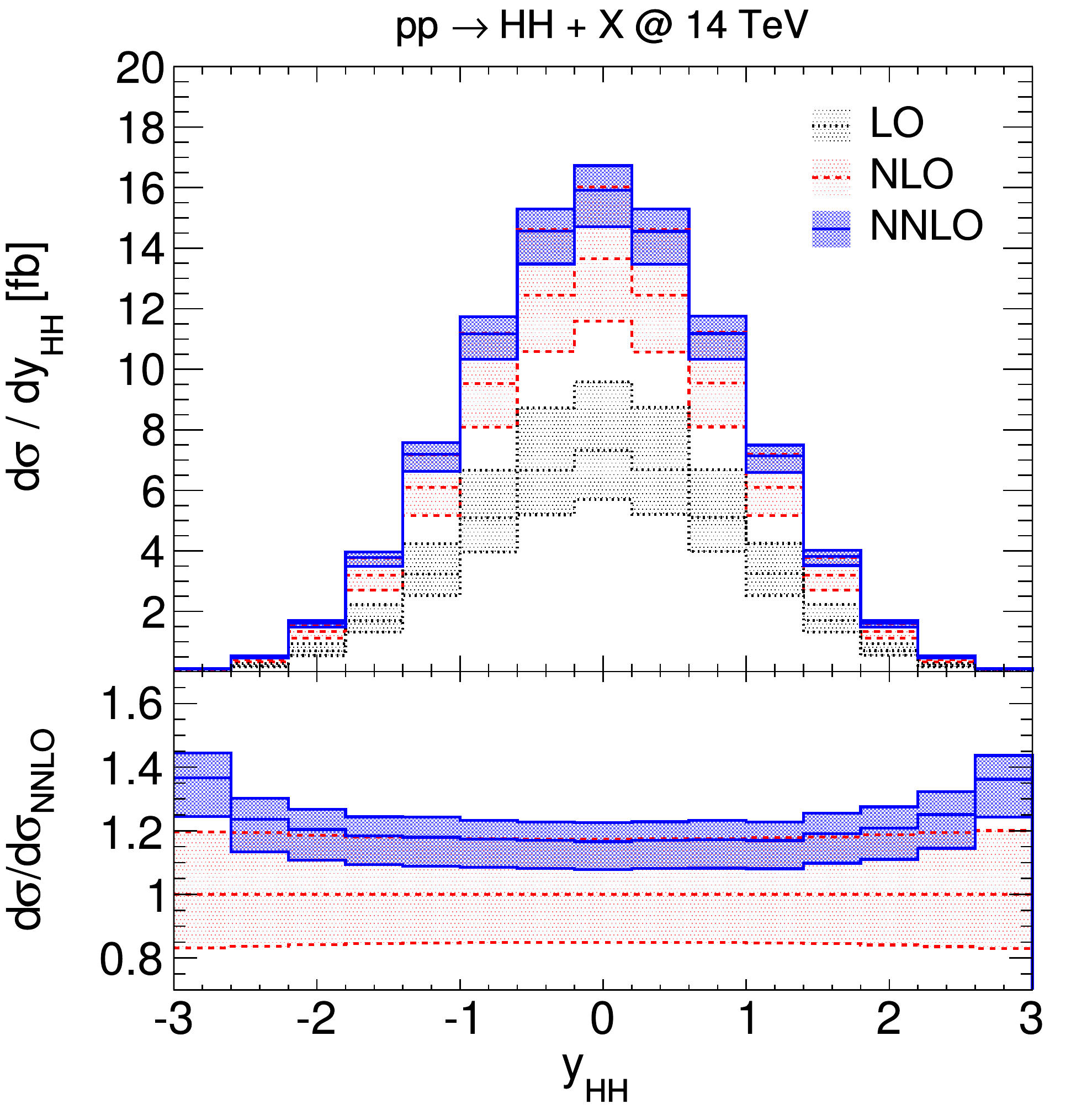}
\caption{
	  Invariant-mass distribution $m_{HH}$ (left) and rapidity distribution $y_{HH}$ (right) 
          of the produced Higgs boson pair.  
	  Curves and bands as in \reffi{fig:pTH1pTH2}. Additionally, in the left plot we show the $m_{HH}$ 
          distribution as obtained with the calculation of \citere{deFlorian:2013jea}.
	}
\label{fig:mHHyHH}
\end{figure*}

Differential distributions in the transverse momentum and the rapidity of the 
two Higgs bosons, ordered by their hardness in $\pT$, are shown in 
\reffi{fig:pTH1pTH2} and \reffi{fig:yH1yH2}, respectively. NNLO corrections 
are overall at the level of $10\%-25\%$ with a rather mild phase-space 
dependence. In particular, in the transverse-momentum distribution of the harder Higgs boson, 
$\pTHone$, the NNLO corrections 
slowly increase as $\pTHone$ increases,
while for the softer Higgs boson the corrections are to a large extent independent of $\pTHtwo$,
except for the very low $\pTHtwo$ region. As for the rapidity distributions, the NNLO effect is largely
constant and at ${\cal O}(20\%)$ for both $y_{H_1}$ and $y_{H_2}$.

In the left plot of \reffi{fig:mHHyHH} we show predictions for the invariant-mass 
distribution of the produced Higgs boson pair, $\mHH$. NNLO corrections 
are at the level of $18\%$ with respect to NLO and hardly show any phase-space dependence.
NNLO predictions in \mHH{} have already been presented in
\citere{deFlorian:2013jea}, and the corresponding results are overlaid in
\reffi{fig:mHHyHH} (left). In the computation of \citere{deFlorian:2013jea} IR
singularities are analytically cancelled, thereby leading to negligible
numerical fluctuations in the shown distribution. Within statistical
uncertainties the results obtained from the two completely independent
implementations agree perfectly.

In the right plot of \reffi{fig:mHHyHH} predictions for the rapidity of the Higgs boson pair, 
$y_{HH}$, are presented. Again, we observe a mild phase-space dependence, with 
increasing NNLO corrections only for large rapidities. 
In all distributions in \reffis{fig:pTH1pTH2}{fig:mHHyHH}, NNLO scale 
uncertainties are reduced to the level of $\pm(5\%-12\%)$, compared to 
$\pm(15\%-20\%)$ at NLO.\\

In \reffi{fig:pTHHpTj1} we show distributions in the transverse momentum of the
Higgs boson pair, $\pTHH$, and of the hardest jet, $\pTj$.
At NLO (which is effectively LO for non-vanishing transverse momenta) 
these two distributions are directly related, 
and in both distributions scale uncertainties reach almost $50\%$. The NNLO effect is
larger on the $\pTHH$ distribution than on the $\pTj$ 
distribution, reaching $80\%$ for $\pTHH\approx 200$\,GeV compared to $60\%$ for 
$\pTj\approx 200$\,GeV.
The NLO nature of these NNLO corrections
is furthermore reflected by sizable scale uncertainties at the level of $30\%-40\%$.
In the limit $\pTHH \to 0$ the perturbative expansion fails due to the appearance
of large logarithmic terms of the form $\log^n\left(\pTHH/\mHH\right)$. Here,
a proper resummation of such terms is required in order to achieve a reliable 
theoretical prediction.

\begin{figure*}[t!]
\centering
   \includegraphics[width=\relplotwidth\textwidth]{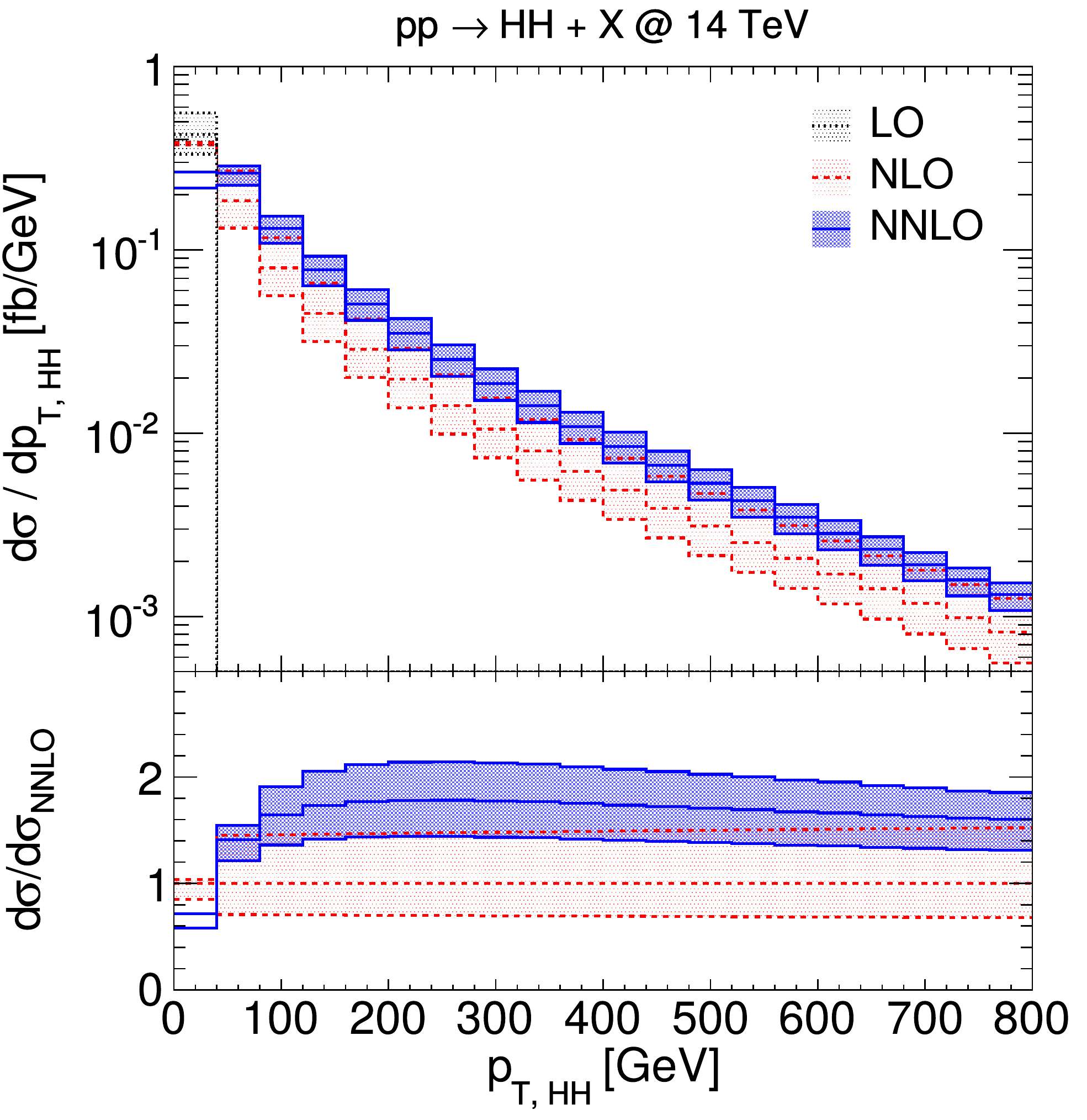}
         \qquad 
   \includegraphics[width=\relplotwidth\textwidth]{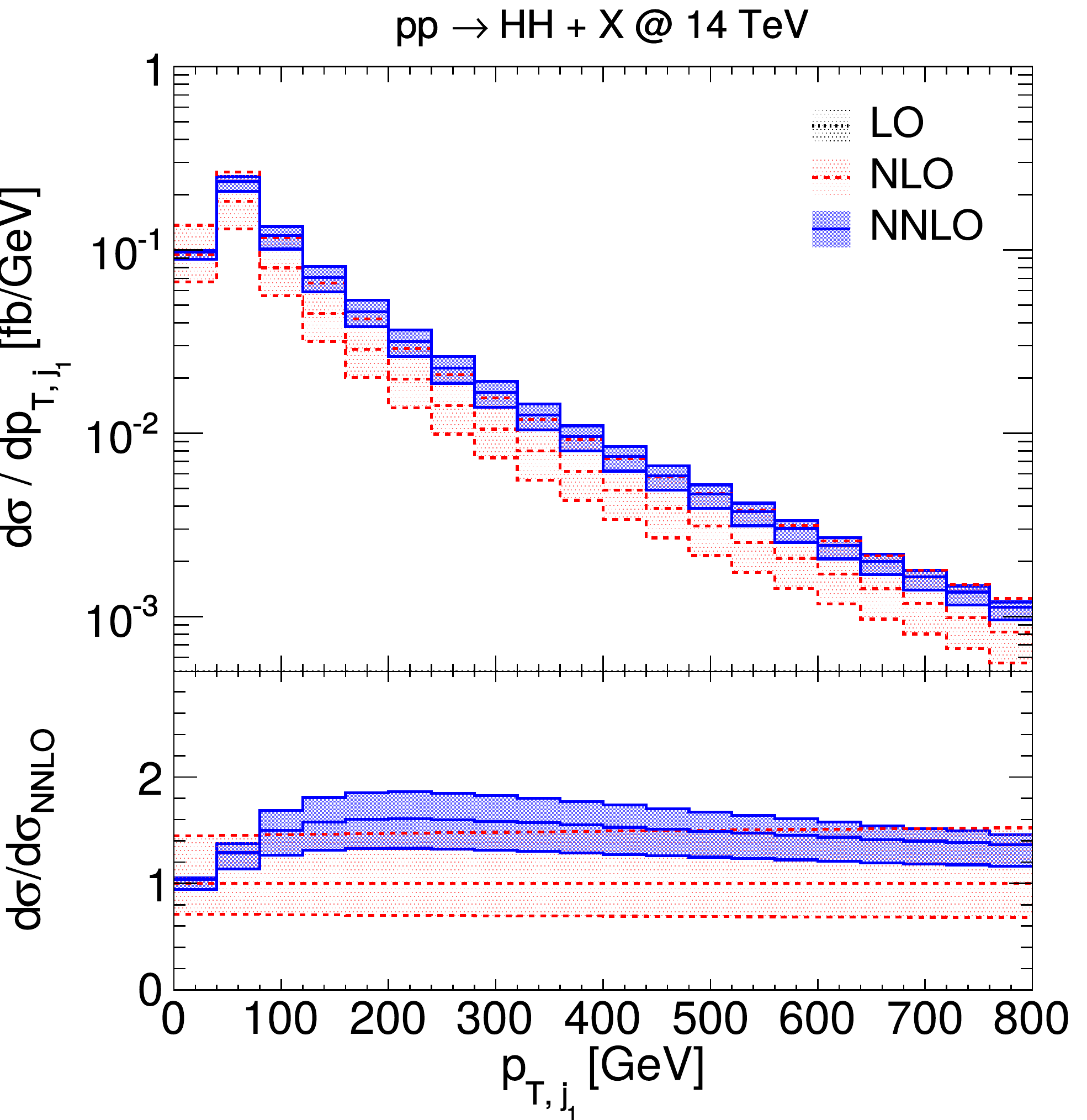}
\caption{
	  Distributions in the transverse momentum of the Higgs boson pair (left) and of the hardest jet (right).
	  Curves and bands as in \reffi{fig:pTH1pTH2}.
	}
\label{fig:pTHHpTj1}
\end{figure*}

\begin{figure*}[t!]
\centering
   \includegraphics[width=\relplotwidth\textwidth]{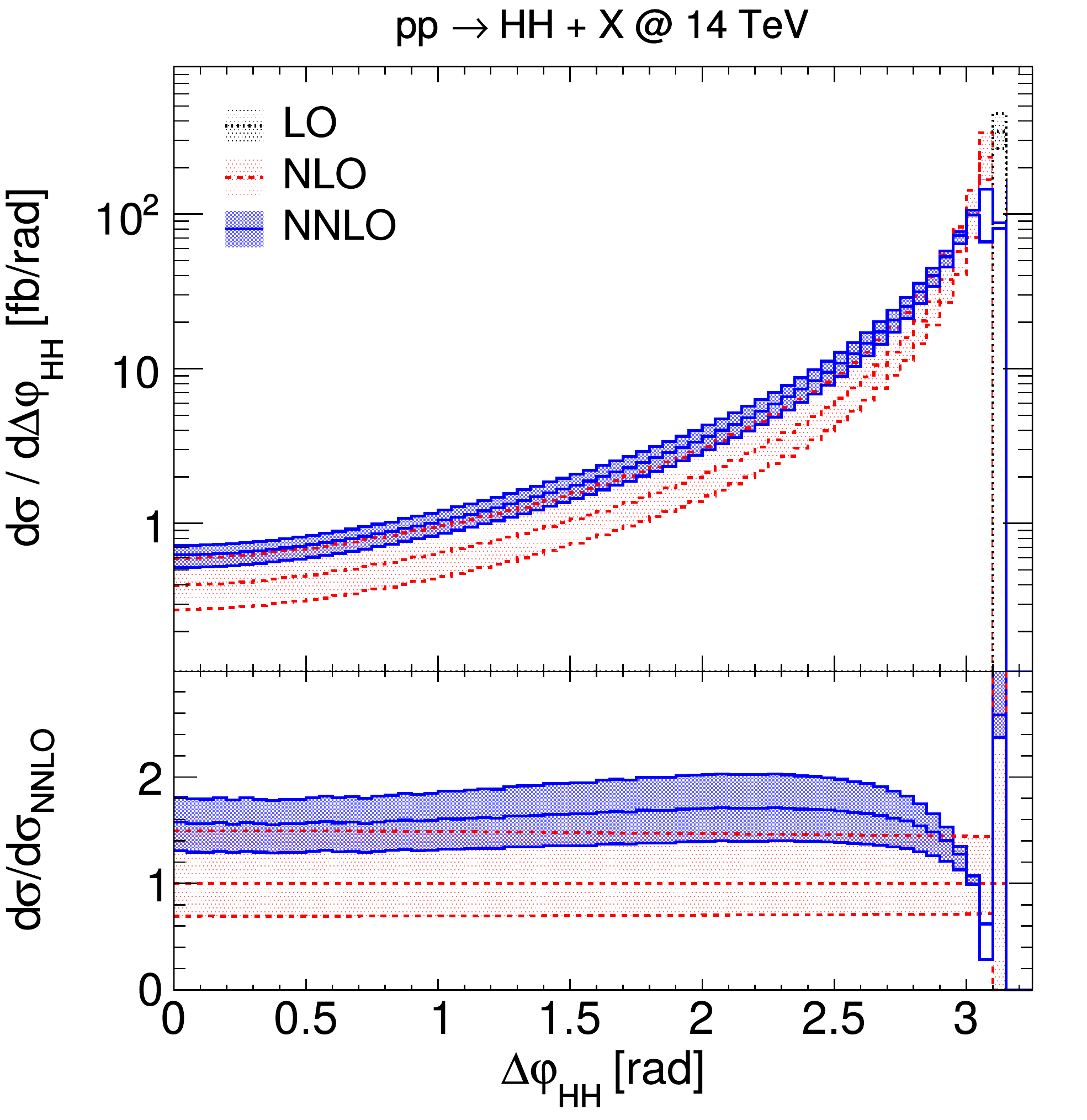}
\caption{
	  Distribution in the angular separation between the two Higgs bosons $\Delta\phi_{HH}$.
	  Curves and bands as in \reffi{fig:pTH1pTH2}.
	}
\label{fig:dPhiHH}
\end{figure*}

At LO the two Higgs bosons are always produced back-to-back. However, at higher 
orders additional QCD radiation allows for a non-trivial angular separation
between the two Higgs bosons. In \reffi{fig:dPhiHH} we show the corresponding 
distribution in the azimuthal angle between the two Higss bosons, $\Delta\phi_{HH}$. 
In our fixed-order approach, NNLO corrections are large and positive in the back-to-back configuration, 
where they are driven by soft-gluon emission, and jump to negative values for $\Delta\phi_{HH}\lesssim \pi$,
due to the mis-cancellation between real and virtual contributions.
In this region of phase-space, large logarithmic terms should again be 
resummed for achieving a reliable theoretical prediction. Configurations at small angles, 
i.e. $\Delta\phi_{HH}\to 0$, are driven by hard gluon emission, and NNLO corrections
are at the level of $60\%$ with respect to NLO.

Finally, in \reffi{fig:dRHj} we investigate corrections to the $\Delta R$ 
separation between the two Higgs bosons and the hardest jet. Overall corrections to 
these observables are moderate at the level of $20\%-40\%$ with largely overlapping 
uncertainty bands between NNLO and NLO. However, for small $\Delta R_{H_1 j_1}$
separations, due to the ordering of the Higgs bosons according to their transverse momenta the
entire phase-space opens up only at the NNLO level, inducing sizable correction
factors at the NLO boundary $\Delta R_{H_1 j_1}\gtrsim \pi/2$.

\begin{figure*}[t!]
\centering
   \includegraphics[width=\relplotwidth\textwidth]{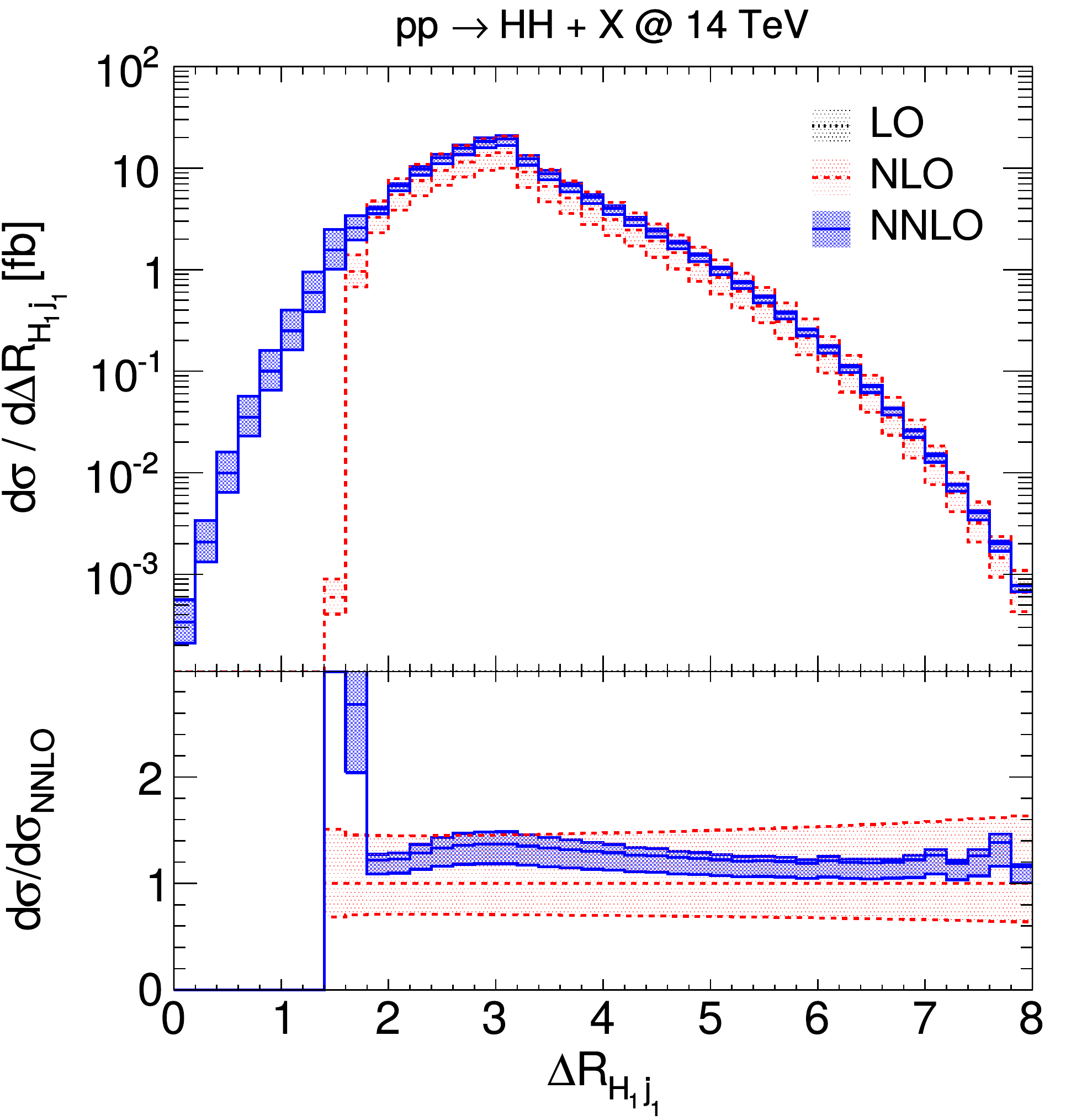}
         \qquad 
   \includegraphics[width=\relplotwidth\textwidth]{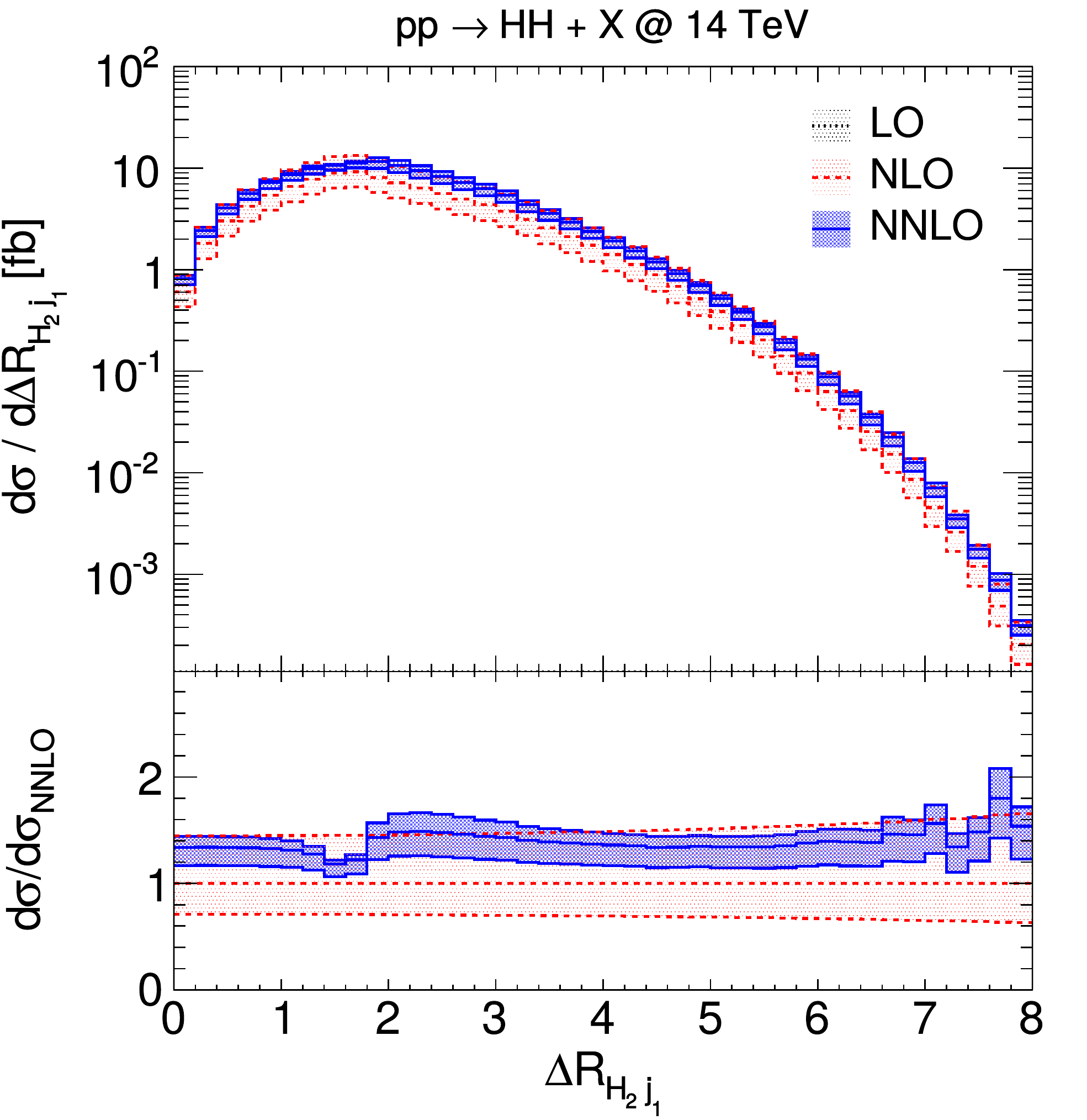}
\caption{
	  Distributions in the $\Delta R=\sqrt{(\Delta\phi)^2+(\Delta y)^2}$ separation between the 
          harder Higgs boson and the hardest jet (left), $\Delta R_{H_1 j_1}$, and the softer Higgs boson
          and the hardest jet (right), $\Delta R_{H_2 j_1}$.
	  Curves and bands as in \reffi{fig:pTH1pTH2}.
	}
\label{fig:dRHj}
\end{figure*}

\section{Summary and Outlook}
\label{sec:summary}

In this paper we have presented the first fully differential calculation for
double Higgs boson production through gluon fusion in hadron collisions.
We worked in the heavy-top limit, and presented results for
differential 
distributions through NNLO QCD for various observables including the transverse-momentum and 
rapidity distributions of the two Higgs bosons. NNLO corrections amount to
about $10\%-25\%$ with respect to the NLO prediction
and are mildly dependent on the kinematics. 
The residual scale uncertainty at NNLO is about $5\%-15\%$.
Only at NNLO the perturbative expansion starts to converge,
and the uncertainty bands obtained through scale variations at NLO and NNLO
overlap in most of the phase-space.
The calculation includes NLO QCD predictions for \pphhj{}. Corrections to the 
corresponding distributions exceed $50\%$ with a residual scale dependence of about
$20\%-30\%$.

The calculation presented here is based on the combination of the $\qT$ subtraction 
formalism~\cite{Catani:2007vq} with the
Monte Carlo framework \Munich,  
supplemented by tree and one-loop amplitudes 
from \OpenLoops~\cite{OLhepforge}. This framework, to be integrated in the
new numerical program \Matrix{}\footnote{\Matrix{} is the abbreviation of 
``\Munich{} Automates qT subtraction and Resummation
to Integrate Cross Sections'', by M.~Grazzini, S.~Kallweit, D.~Rathlev, M.~Wiesemann. In preparation.}, 
which is currently under development, allows for an extremely flexible implementation.

In the present paper we have limited ourselves to strictly work in the 
heavy-top limit of the SM.
With the exact NLO virtual contributions available since recently, 
a combination of the exact results at NLO accuracy with the 
NNLO calculation in the heavy-top limit should be performed in the future.
This combination, together with the inclusion of the Higgs boson decays,
will facilitate realistic phenomenological studies at an unprecedented level
of precision, as required for future measurements of the Higgs trilinear coupling.


\noindent {\bf Acknowledgements.}
We thank Stefano Pozzorini and Marius Wiesemann for valuable discussions. 
This research was supported in part by the Swiss National Science 
Foundation (SNF) under contracts CRSII2-141847, 200021-156585, 
and by the Research Executive Agency (REA) of the European 
Union under the Grant Agreement number PITN--GA--2012--316704 ({\it HiggsTools}).

\bibliographystyle{UTPstyle}
\bibliography{hhnnlo}

\end{document}